\documentclass[10pt,a4paper]{article}
\usepackage[a4paper,left=2.5cm,right=2.5cm,top=2.5cm,bottom=2.5cm]{geometry}
\usepackage[super,compress]{cite}
\usepackage{graphicx}
\usepackage{authblk}

\usepackage{graphicx,latexsym} 
\usepackage{amssymb,amsmath}
\usepackage{amsthm}

\usepackage{hyperref}

\begin{document}

\title{On attractor behavior in braneworld constant-roll inflation}

%\author{N.B, D.D., G.Dj., M.M., M.S.}

\author[1]{Goran S. Djordjevic\thanks{gorandj@junis.ni.ac.rs}}
\author[2]{Neven Bili\'c\thanks{bilic@irb.hr}}
\author[1]{Dragoljub D. Dimitrijevic\thanks{ddrag@pmf.ni.ac.rs}}
\author[1]{Milan Milosevic\thanks{milan.milosevic@pmf.edu.rs}}
\author[3]{Marko Stojanovic\thanks{marko.stojanovic@pmf.edu.rs}}

\affil[1]{Department of Physics, University of Ni\v s, Serbia}
\affil[2]{Division of Theoretical Physics, Rudjer Bo\v{s}kovi\'{c} Institute, Zagreb, Croatia}
\affil[3]{Faculty of Medicine, University of Ni\v s, Serbia}

\maketitle

\begin{abstract}
We investigate in detail the attractor behavior of some inflationary models based on braneworld dynamics 
under the constant-roll condition. We describe the dynamics of the models, assuming that the second slow-roll parameter remains constant during inflation. 
We show that the dynamics of the considered models have the property of a cosmological attractor. 
\end{abstract}

%\tableofcontents


\section{Introduction}

The inflation hypothesis proposes an accelerated expansion of the universe in a very short time to solve some problems in the hot Big Bang model [\citen{Guth:1980zm}]. Besides, inflation provides the seed for the large-scale structure of the universe. The most common way to realize inflationary expansion is to introduce one single scalar field, known as the inflaton field, associated with the unknown form of matter with negative pressure. The most straightforward realization of the inflationary stage can be achieved via a scalar field with a kinetic term and a nearly flat potential. A prospective candidate for inflaton is the tachyon scalar field. The primary motivation for using the tachyon field in cosmological models is based on the fact that matter with a negative pressure is produced during tachyonic condensation. This process originates in string theory and is described by the effective Lagrangian in Dirac-Born-Infeld (DBI) form, with a non-canonical kinetic term [\citen{Sen:1999md,Feinstein:2002aj}]. In addition, the tachyon field could contribute to dark matter at late times [\citen{Sen:2002nu}]. 

Intuitively, the attractor behavior of the inflationary scenario is a necessary condition for a successful model of inflation. It guarantees a unique evolution of the universe, independent of initial conditions. In the theory of dynamical systems, an attractor behavior describes situations where a collection of phase-space points evolve into a specific region and never leave. A more formal definition is given in [\citen{Steven H. Strogatz}]. Although it was shown that this notion in cosmological models can not be well defined, it makes sense to introduce the effective phase space where attractor-like behavior may appear [\citen{Remmen:2013eja}]. 

The inflationary scenario, named the slow-roll inflation, is based on the idea that the inflaton slowly rolls down from the top of its potential. Extensive studies of inflationary scenarios based on a new idea that one of the slow-roll parameters is constant during inflation, have been carried out in recent years [\citen{Motohashi:2014ppa,Anguelova:2017djf,Yi:2017mxs,Gao:2018tdb}]. In this work, we aim to investigate for the first time the attractor behavior in models of tachyon inflation constructed in the framework of the second Randall-Sundrum (RSII) type of braneworld [\citen{Randall:1999vf}], assuming that the second slow-roll parameter $\eta$ is being constant. 

The paper is organised as follows. In Section 2 we describe the properties of tachyon dynamics in inflationary models. In Section 3, we introduce a scenario of the constant-roll inflation. Then, we apply this formalism to some models governed by the braneworld dynamics, assuming that the inflaton is a tachyon field. In Section 4, we discuss the attractor behavior in constant-roll models with a constant parameter $\eta$. The conclusions are given in Section 5.

\section{Tachyon dynamics in inflationary models}

During inflation, the universe is assumed to be isotropic and homogeneous, described by the spatially flat FLRW metric 
\begin{equation}
	ds^2=-dt^2+a^2(t)\delta_{ij}dx^{i}dx^{j},
\end{equation}
where  $a(t)$ is the scale factor that describes the universe expansion. 

The dynamics of the tachyon field $\theta$ in the case of space homogenity and isotropy is obtained from the  Lagrangian of the DBI form [\citen{garousi,Steer:2003yu}]
\begin{equation}
	{\cal L}=-V(\theta)\sqrt{1-\dot{\theta}^2},
\end{equation}
where the dot denotes a derivative with respect to time. The tachyon potential $V(\theta)$ has a global maximum at $\theta=0$ and minima as $|\theta|\rightarrow\infty$ where $V\rightarrow 0$. 

The energy density and pressure of the ideal fluid associated with the tachyon  field are given by 
\begin{equation}
	\rho=\frac{V}{\sqrt{1-\dot{\theta}^2}}, \label{eqrho}  
\end{equation}
\begin{equation}
	p=-V\sqrt{1-\dot{\theta}^2}.
	\label{eqp} 
\end{equation}
The energy-momentum conservation 
\begin{equation}
	\dot{\rho}-3H(\rho+p)=0,
\end{equation}
gives the equation of motion for the tachyon field 
\begin{equation}
	\frac{\ddot{\theta}}{1-\dot{\theta}}+3H\dot{\theta}+\frac{V_{,\theta}}{V}=0,
	\label{eqtheta}
\end{equation}
where the subscript ${,}\theta$ denotes a derivative with respect to $\theta$ and the Hubble parameter $H(t)$ is the function of the scale factor 
\begin{equation}
	H(t)=\frac{\dot{a}}{a}.
\end{equation}
As we will see,  equation (\ref{eqtheta}) will be used to study the attractor behavior of the model. 

\section{Constant-roll inflation}

The system of dynamical nonlinear equations in an inflationary model usually can not be solved analytically, except for some particular type of the potential. The most common approach to finding solutions is based on slow-roll approximation, assuming that the Hubble slow-roll parameters [\citen{Liddle:1994dx}]
\begin{equation}
	\epsilon=-\frac{{\dot{H}}}{H^2} ,
	\label{epsilon}
\end{equation} 
and
\begin{equation}
	\eta=-\frac{\ddot{H}}{2H\dot{H}} ,
	\label{defeta1}
\end{equation}
are much smaller than unity at the beginning of inflation. The slow-roll parameters (\ref{epsilon}) and (\ref{defeta1}) were initially introduced in a model with a single canonical scalar field $\phi$. The action for the model is given by 
\begin{equation}
	S=\int d^4x\sqrt{-g}\left(\frac{1}{2M_{\rm Pl}^{2}}R-\frac{1}{2}g^{\mu\nu}\partial_{\mu}\phi\partial_{\nu}\phi-V(\phi)\right),
	\label{actionphi}
\end{equation}
where $M_{\rm Pl}=(8\pi G_{\rm N})^{-1/2}$ (with $G_{\rm N}$ being Newton's gravitational constant) and $V(\phi)$ is a potential of the scalar field $\phi$. The Friedmann equations for a spatially flat FLRW metric are of the form 
\begin{eqnarray}
	H^2&=&\frac{1}{3 M_{\rm Pl}^2}\left(\frac{1}{2}\dot{\phi}^2+V(\phi)\right),
	\label{F1scalar}\\
	\dot{H}&=&-\frac{1}{2M_{\rm Pl}^2}\dot{\phi}^2
	\label{F2scalar}.
\end{eqnarray}
The equation of motion  for the scalar field $\phi$, obtained from the action (\ref{actionphi}), is of the form 
\begin{equation}
	\ddot{\phi}+3H\dot{\phi}+V_{,\phi}=0,
	\label{EMphi}
\end{equation}
and the parameter $\eta$ becomes
\begin{equation}
	\eta=-\frac{\ddot{\phi}}{H\dot{\phi}}.
	\label{eta} 
\end{equation}
It can be shown that the slow-roll approximation is satisfied if the kinetic energy of the inflaton field is negligible compared with the potential energy, 
i.e., $1/2\dot{\phi}^2\ll V(\phi)$. In the slow-roll approximation, the first term of the right-hand side of (\ref{EMphi}) can be neglected ($\ddot{\phi}\simeq 0$), yielding to $\eta\simeq 0$ and
\begin{equation}
	3H\dot{\phi}+V_{,\phi}\simeq 0.
\end{equation}
In some cases, for instance, when the potential of the scalar field is exactly flat ($V_{,\phi}= 0$), instead of the slow-roll regime the   ultra slow-roll regime occurs [\citen{Martin:2012pe}]
\begin{equation}
	\ddot{\phi}+3H\dot{\phi}= 0,
\end{equation}
corresponding to $\eta= -3$. The ultra slow-roll inflation is generalized to constant-roll inflation [\citen{Motohashi:2014ppa,Anguelova:2017djf}], the regime during which the second slow-roll parameter $\eta$, defined by (\ref{eta}), is constant and could be of the order of unity. In this class of models, defining the potential is unnecessary, which is an important advantage. 
Besides, the dynamical system of equations often has analytical solutions [\citen{Yi:2017mxs}].   

More generally, slow-roll inflation can be defined by demanding that one of the slow-roll parameters is constant during inflation, using various definitions of slow-roll parameters [\citen{Gao:2018tdb}]. In this work, we will discuss the constant-roll models with parameter $\eta$ given by expressions (\ref{defeta1}) and (\ref{eta}). 

\subsection{Tachyon constant-roll inflation in RSII cosmology}

In this section, we review the main properties of the tachyon constant-roll inflation model in RSII cosmology with a constant parameter $\eta$. This model is discussed in detail in reference [\citen{Stojanovic:2023qgm}].

The RSII model describes a universe containing two branes with opposite tensions embedded in a 4+1-dim asymptotically Anti-de Sitter (AdS\textsubscript{5}) space, observers reside on the positive tension brane, and the negative tension brane is pushed off to infinity [\citen{Randall:1999vf}]. The branes are separated in the fifth dimension, where only gravity can propagate. The dynamics of the Hubble parameters in RSII cosmology is described by the Friedmann equations, which, in the case of a flat FLRW universe, are of the form [\citen{Mohammadi:2020ftb}]
\begin{equation}
	H^{2}=\frac{8\pi}{3M_{4}^2}\rho\left(1+\frac{\rho}{2\lambda}\right),
	\label{F1RSII}
\end{equation}
\begin{equation}
	\dot{H}=-\frac{4\pi}{M_{4}^2}(\rho+p)\left(1+\frac{\rho}{\lambda}\right).
	\label{F2RSII}
\end{equation}
Unlike the standard (four-dimensional) cosmology, the Friedmann equations (\ref{F1RSII}) and (\ref{F2RSII}) have the additional terms in the function of the brane tension $\lambda$, which is related to the five-dimensional and four-dimensional Planck masses,
\begin{equation}
	\lambda=\frac{3}{4\pi}\left(\frac{M_{5}^3}{M_{4}}\right)^2,
\end{equation}
where $M_{4}=G_{\rm N}^{-1/2}$. Following reference [\citen{Mohammadi:2020ftb}], we consider the case when the energy density is much larger than the brane tension, i.e., for $\rho\gg\lambda$. As a consequence, equations (\ref{F1RSII}) and (\ref{F2RSII}) take a simplified form
\begin{equation}
	H^{2}\simeq\frac{4\pi}{3M_{4}^2}\frac{\rho^2}{\lambda},
	\label{F1approx}
\end{equation}
\begin{equation}
	\dot{H}\simeq -\frac{4\pi}{M_{4}^2}\frac{\rho}{\lambda}.
	\label{F2approx}
\end{equation}
According to the Hamilton-Jacobi formalism, we can consider the Hubble parameter as a function of the tachyon field $H=H(\theta)$ and express the time derivative of $H$ via $\dot{H}=H_{,\theta}\dot{\theta}$. Then, combining Friedman’s equations, (\ref{F1approx}) and (\ref{F2approx}), with (\ref{eqrho}) and (\ref{eqp}) one obtain   
\begin{equation}
	\dot{\theta}=-\frac{1}{3}\frac{H_{,\theta}}{H^{2}}.
	\label{dottheta2} 
\end{equation}
The expression for $\dot{\theta}$ differs from the corresponding expression in the standard tachyon cosmology [\citen{Mohammadi:2018oku}],  where the value of $\dot{\theta}$ is lower by a factor of 2. 

For constant $\eta$, the expression (\ref{defeta1}) besomes a differential equation for the Hubble parameter
\begin{equation}
	\ddot{H}+2\eta H\dot{H}=0.
	\label{eq22}
\end{equation}
Using (\ref{dottheta2}), equation (\ref{eq22}) can be transformed to a differential equation with respect to $\theta$ 
\begin{equation}
	H_{,\theta\theta}H-H_{,\theta}^2-3\eta H^{4}=0,
	\label{difH}
\end{equation}
with solution 
\begin{equation}
	H=\frac{2 e^{\theta}}{e^{2\theta}-3\eta}.
	\label{Htheta}
\end{equation}

It was shown in reference [\citen{Stojanovic:2023qgm}] that the constant-roll model with a constant parameter $\eta$ in RSII cosmology gives a similar prediction for the observational parameters as the equivalent model in standard cosmology. A good agreement with the observational data [\citen{Planck:2018jri}] is obtained for negative and small $\eta$. In that model it has been proven [16] that different cosmologies do not affect the results for observation parameters at least at linear order of the slow roll parameters. 
Given that result, it would be worthwhile to apply the constant-roll formalism to the holographic RSII model with another definition for the second slow-roll parameter.

\subsection{Tachyon constant-roll inflation in holographic cosmology} 

The holographic RSII model is based on a scenario in which the 4-dim brane is located at the holographic boundary of a 4+1-dim asymptotically AdS\textsubscript{5} space. The effective four-dimensional Einstein equations on the holographic boundary yield  holographic Friedmann equations [\citen{Bilic:2018uqx}]
\begin{equation}
	h^2-\frac{1}{4}h^4=\frac{\kappa^2}{3}\ell^4\rho,\label{F1}
\end{equation}
\begin{equation}
	\dot{h}\left(1-\frac{1}{2}h^2\right)=-\frac{\kappa^2}{2} \ell^3(p+\rho),\label{F2}
\end{equation}
where $\ell$ is AdS\textsubscript{5} curvature radius,  $h\equiv \ell H$ is a dimensionless Hubble parameter and $\kappa$ is fundamental coupling parameter 
defined by 
\begin{equation}
	\kappa^{2}=\frac{8\pi G_{\rm 4}}{\ell^2}.
\end{equation}
The tachyon field $\theta(t)$ and the time $t$ are measured in units of $\ell$.

Suppose the parameter $\eta$, defined by (\ref{eta}), is constant during inflation. A similar assumption was made in standard cosmology [\citen{Mohammadi:2018oku}]. Combining holographic Friedmann's equations, (\ref{F1}) and (\ref{F2}), with the expressions for energy density and pressure of tachyons field, (\ref{eqrho}) and (\ref{eqp}), one finds
\begin{equation}
	\dot{\theta}=-\frac{2\ell}{3}\frac{h,_{\theta}}{h^2}\frac{1-\frac{1}{2}h^2}{1-\frac{1}{4}h^2},
	\label{dottheta}
\end{equation}
where, according to the Hamilton-Jacobi formalism, the expansion rate $h$  is a function of the tachyon field $\theta$. Taking the time derivative of (\ref{dottheta})  and replacing $\dot{\theta}$ and $\ddot{\theta}$ in (\ref{eta}) we get a differential equation for the Hubble parameter 
\begin{equation}
	hh_{,\theta\theta}-2h,_{\theta}^2\left(1+\frac{h^2}{4(1-\frac{1}{2}h^2)(1-\frac{1}{4}h^2)}\right)+\frac{3}{2\ell^2}\eta h^4\frac{1-\frac{1}{4}h^2}{1-\frac{1}{2}h^2}=0.\label{hthetatheta}
\end{equation}
The obtained equation is more complicated than the corresponding equation in the model from standard cosmology and will be solved numerically.

\section{Attractor behaviour}

In this section, we demonstrate the attractor behavior of the inflationary solution. The naturalness in cosmic inflation suggests the existence of cosmological attractors: the dynamical conditions under which evolving scalar fields approach a unique behavior without fine-tuning of initial conditions [\citen{Remmen:2013eja}]. The definition of attractor in cosmic inflation is less rigorous than the definition in the dynamical system theory. In this context, it is useful to recall that the condition for the attractor behavior  was considered in [\citen{Mukhanov:2005sc}], in the form $d\dot{\phi}/d\phi\approx 0$. The futility of strictly mathematically defining the attractor in an inflationary model is indicated in reference [\citen{Remmen:2013eja}].

The attractor behavior is a necessary condition for a successful model of inflation. To find a numerical solution for the system of dynamical equations in an inflationary model, one must choose initial conditions. In an inflationary model, the potential usually has one free parameter, corresponding to the freedom of specifying the Hubble parameter at some initial time. As a consequence,  this freedom leads to more solutions to dynamical equations. If the attractor solution exists, the solutions corresponding to a wide range of initial conditions will converge to the same inflationary stage. Hence, one of these solutions for chosen initial conditions is a good representative of all the solutions.  The importance and the properties of the inflationary attractor solution in the slow-roll regime were discussed in reference [\citen{Liddle:1994dx}] while the attractor behavior in inflation model with tachyon field in standard cosmology and RSII cosmology was initially discussed in reference [\citen{Guo:2003nz}]. Besides, the dynamical equations in the slow-roll approximation,  become first-order differential equations. In other words, the initial value of $\dot{\theta}$ can be taken arbitrarily,  which makes sense only if there is an attractor behavior. 

As pointed out in reference [\citen{Mohammadi:2022vru}], there are two ways of studying the attractor behavior of the inflationary models: analytically or numerically. The analytical procedure is customary for the inflationary model with the Hamilton-Jacobi approach. Following this method, substituting the perturbed Hubble parameter, i.e., $H_{0}+\delta H$, in the Hamilton-Jacobi equation, one obtains an equation for the perturbation  $\delta H$. If the perturbation decays, the solution $H_{0}$ will be stable and will indicate the attractor properties. To prove the attractor behavior numerically, it is necessary to solve the equation of motion (in our models equation (\ref{eqtheta})) and study the properties of the solution in an effective phase space with coordinates $\theta$ and $\dot{\theta}$. This procedure is possible only if the function $H(\theta)$ is known. The function $H(\theta)$ can be obtained by solving a system of dynamical equations in the model.

It is well known that the model of inflation with a canonical scalar field has the properties of the cosmological attractor.  To demonstrate these properties, let us analyze the solution to the equation of motion for $V(\phi)=m^2\phi^2/2$, where $m$ is the mass of the field $\phi$.  With the help of (\ref{F1scalar}), equation (\ref{EMphi}) can be expressed in the following form
\begin{equation}
	\ddot{\phi}+\dot{\phi}\sqrt{\frac{3}{2 M_{\rm Pl}^2}\left(\dot{\phi}^2+m^2\phi^2\right)}+m^2\phi=0.
	\label{eqforphi}
\end{equation}
This equation is a nonlinear second-order differential equation for a damped harmonic oscillator, where the damping coefficient is a function of $\phi$ and $\dot{\phi}$.  
\begin{figure}[h]
	\begin{center}
		\includegraphics[scale=0.8]{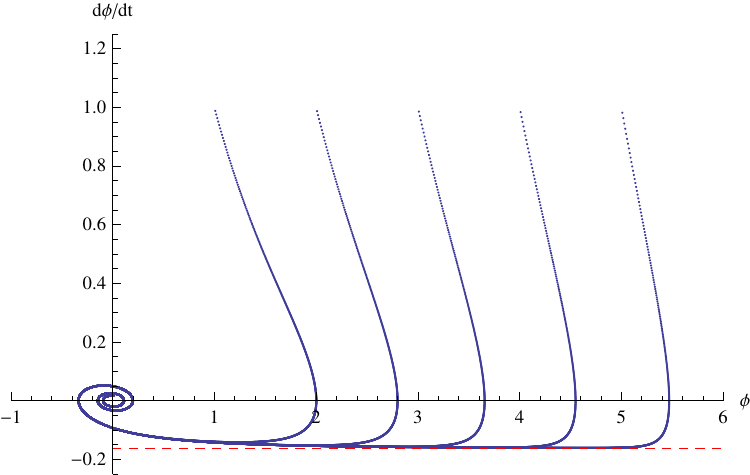}
		\caption{Phase portrait of the canonical field with potential $V(\phi)=m^2\phi^2/2$, for $m=0.2M_{\rm Pl}$, in units where $M_{\rm Pl}=1$ [\citen{Remmen:2013eja}].}
		\label{attractorsc}
	\end{center}
\end{figure}
In Fig. \ref{attractorsc}, we present the numerical solution of the equation (\ref{eqforphi}) in the effective phase space. Solid lines represent the numerical solutions to the equation (\ref{eqforphi}) obtained for random initial conditions. For large field values, the solution represented by a dashed line approaches the horizontal line
\begin{equation}
	\dot{\phi}=-\sqrt{\frac{2}{3}}m,
\end{equation}
which attracts all other trajectories.  

Cosmological attractor solutions were also found in the tachyon model in standard and RSII cosmology for the exponential type of potential [\citen{Guo:2003nz,Guo:2003zf}].

\subsection{Attractor behavior in RSII cosmology}

\begin{figure}[h]
\begin{center}
\includegraphics[scale=0.8]{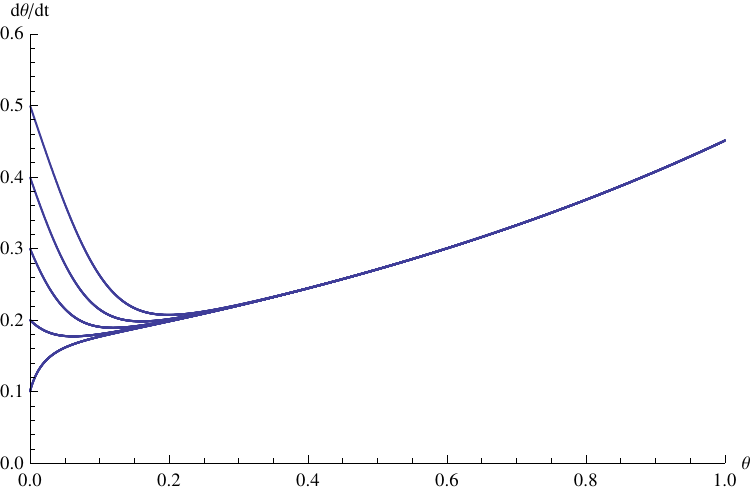}
\caption{The phase space trajectories obtained by solving the equation (\ref{eqtheta}) numerically for $\theta_{\rm i}=0$ and $\dot{\theta}_{\rm i}$ in range $0.2\leq\dot{\theta}_{\rm i}\leq0.5$, for the model with $\eta=-0.013$ in units where $3M_{5}^{3}/(4\pi)=1$.}
\label{phase space}
\end{center}
\end{figure}

As explained in the previous section, to examine the attractor behavior of our models numerically,  we need to specify the potential. 
The potential $V(\theta)$ which provides the evolution of the Hubble expansion rate given by (\ref{Htheta}) 
can be reconstructed using (\ref{eqrho}) and (\ref{dottheta2}) together with (\ref{F1approx}). We obtain
\begin{equation}
V(\theta)=\frac{3M_{5}^{3}}{4\pi}H\sqrt{1-\frac{1}{9}\frac{H_{,\theta}^{2}}{H^{4}}}=\frac{3M_{5}^{3}}{4\pi}\sqrt{\frac{36e^{2\theta}-(e^{2\theta}+3\eta)^2}{9(e^{2\theta}-3\eta)}}.
\label{V}
\end{equation}
Substituting (\ref{Htheta}) and (\ref{V}) in (\ref{eqtheta}) we obtain the differential equation for $\theta$. The numerical solutions obtained for arbitrary initial values $\dot{\theta}_{\rm i}$ and $\theta_{\rm i}$ at $t=0$, are displayed in ($\dot{\theta}$, $\theta$) plane (Fig. \ref{phase space}).  
We conclude that there is a curve that attracts all trajectories obtained for any initial conditions, 
which proves that the inflationary trajectories are attractors. 
We can also see that the dynamical system quickly reaches the attractor points, providing the unique inflationary phase. 
Since the parameter $\eta$ must be very small to provide a good agreement of the model with observational data [\citen{Stojanovic:2023qgm}], 
the obtained phase portrait is very similar to the phase portrait of the standard tachyonic slow-roll inflation [\citen{Guo:2003zf}].

\subsection{Attractor behavior in holographic cosmology}

\begin{figure}[h]
	\begin{center}
		\includegraphics[scale=0.6]{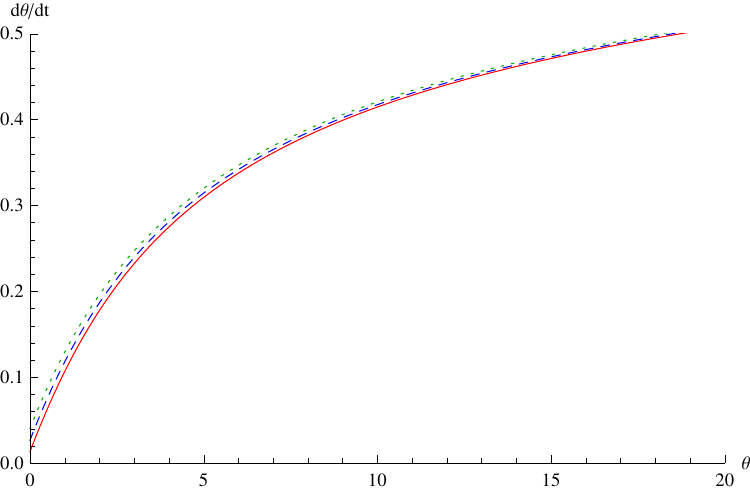}
		\hspace{0.5cm}
		\includegraphics[scale=0.6]{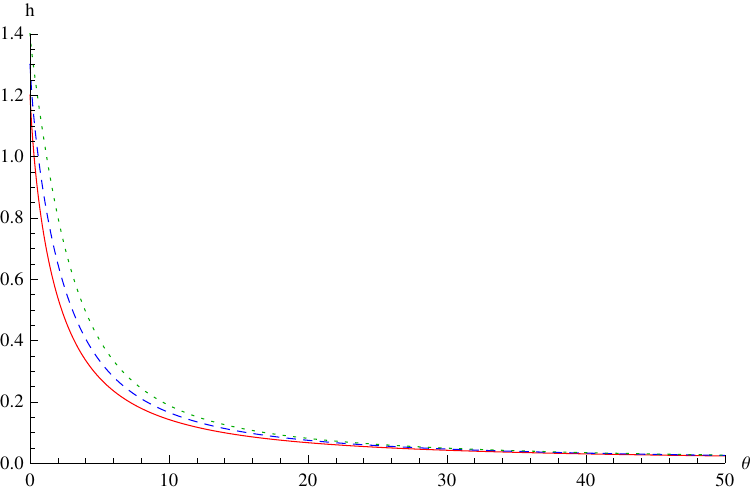}
		\caption{The phase space trajectories (left panel) obtained by solving equation (\ref{hthetatheta}) numerically for $h_{\rm i}=1.4$ and various $h_{,\theta_{\rm i}}$ in interval $-0.1\leq h_{,\theta_{\rm i}}\leq -0.3$, and the solutions to equation (\ref{hthetatheta}) (right panel) for $h_{,\theta_{\rm i}}=-0.1$ and various $h_{\rm i}$ in interval $1.2\leq h_{\rm i}\leq  1.4$, for the model with $\eta=0.077$. }
		\label{phasespace}
	\end{center}
\end{figure}

In the holographic model, the potential may be reconstructed as it was shown in the previous subsection. We obtain
\begin{equation}
	V=\frac{3}{\kappa^2}h^2(1-\frac{1}{4}h^2)\sqrt{1-\frac{4\ell^2}{9}\frac{h_{,\theta}^2}{h^4}\left(\frac{1-\frac{1}{2}h^2}{1-\frac{1}{4}h^2}\right)^2}.
	\label{potentialhol}
\end{equation}
To analyze the attractor behavior, in this case instead of equation (\ref{eqtheta}) we have to solve equation (\ref{hthetatheta}) numerically for some chosen initial values of  $h$ and $h_{,\theta}$ for a fixed value of the parameter $\eta$. The expression (\ref{dottheta}) may be integrated numerically, yielding the time $t$ as a function of $\theta$. The initial value of $\theta$ can be set to zero and the initial value of $\dot{\theta}$ is fixed by (\ref{dottheta}). The solutions in phase space plotted in  Fig. \ref{phasespace} show a typical attractor behavior. 
\section{Conclusion}

We have studied the attractor behavior of RSII (non) holographic cosmology in constant-roll infation. The considered models, with nonlinear dynamics, are based on the braneworld inflation scenario with a tachyon field.

As our main result, 
by solving the equation of motion for the tachyon field in a braneworld constant-roll scenario with a constant parameter $\eta$ defined by (\ref{defeta1}),
we have shown the existence of attractor solutions. These solutions are completely determined by the evolution of the Hubble parameter $H$. 

In addition, we have proven the attractor behavior in the constant-roll model with the constant parameter $\eta$ defined by (\ref{eta}) in the framework of the holographic cosmology. 

The obtained results confirm the validity of studying inflation in these models. These results also indicate the attractor behavior in more general framework of the braneworld scenario. 

In this context, it would be of considerable interest to investigate if other constant-roll models currently on the market satisfy such attractor features. A straightforward extension of our work would be to study the constant-roll models considered in this paper by applying the dynamical system theory. The presented ideas are the topics of ongoing research.

\section{Acknowledgement}
This work has been supported by the ICTP-SEENET-MTP project NT-03 TECOM-GRASP (ThEoretical and Computational Methods in GRavitation and AStroPhysics) and the COST Action CA18108 "Quantum gravity phenomenology in the multimessenger approach". M. Stojanovic acknowledges the support provided by The Ministry of Science, Technological Development and Innovation of the Republic of Serbia under contract 451-03-65/2024-03/200113. D. D. Dimitrijevic, G. S. Djordjevic, and M. Milosevic acknowledge the support provided by the same Ministry, contract 451-03-65/2024-03/200124. In addition, G. S. Djordjevic acknowledges the support of the CEEPUS Program M-RS-1514-2324-181517 "Gravitation and Cosmology" as well as the hospitality of colleagues at the University of Banja Luka and CERN-TH. N. Bili\' c acknowledges the hospitality of the Department of Physics at the University of Ni\v s, where part of his work was completed.

%\begin{thebibliography}{000} %for 3 digits


\begin{thebibliography}{00}  %for 2 digits
%\begin{thebibliography}{0}    %for 1 digit

%\cite{Guth:1980zm}
\bibitem{Guth:1980zm}
A.~H.~Guth,
% "The Inflationary Universe: A Possible Solution to the Horizon and Flatness Problems,"
Phys. Rev. D \textbf{23} (1981), 347-356
doi:10.1103/PhysRevD.23.347

%\cite{Sen:1999md}
\bibitem{Sen:1999md}
A.~Sen,
% "Supersymmetric world volume action for nonBPS D-branes,"
JHEP \textbf{10}, 008 (1999)
doi:10.1088/1126-6708/1999/10/008
[arXiv:hep-th/9909062 [hep-th]].

%\cite{Feinstein:2002aj}
\bibitem{Feinstein:2002aj}
A.~Feinstein,
% "Power law inflation from the rolling tachyon,"
Phys. Rev. D \textbf{66} (2002), 063511
doi:10.1103/PhysRevD.66.063511
[arXiv:hep-th/0204140 [hep-th]].

%\cite{Sen:2002nu}
\bibitem{Sen:2002nu}
A.~Sen,
% "Rolling tachyon,"
JHEP \textbf{04}, 048 (2002)
doi:10.1088/1126-6708/2002/04/048
[arXiv:hep-th/0203211 [hep-th]].

%\cite{Steven H. Strogatz}
\bibitem{Steven H. Strogatz}
S. H. Strogatz,
Nonlinear Dynamics and Chaos With Applications to Physics, Biology, Chemistry, and Engineering,
CRC Press, 2018,
ISBN 978-0-8133-4910-7

%\cite{Remmen:2013eja}
\bibitem{Remmen:2013eja}
G.~N.~Remmen and S.~M.~Carroll,
% "Attractor Solutions in Scalar-Field Cosmology,"
Phys. Rev. D \textbf{88}, 083518 (2013)
doi:10.1103/PhysRevD.88.083518
[arXiv:1309.2611 [gr-qc]].


%\cite{Motohashi:2014ppa}
\bibitem{Motohashi:2014ppa}
H.~Motohashi, A.~A.~Starobinsky and J.~Yokoyama,
% "Inflation with a constant rate of roll,"
JCAP \textbf{09}, 018 (2015)
doi:10.1088/1475-7516/2015/09/018
[arXiv:1411.5021 [astro-ph.CO]].

%\cite{Anguelova:2017djf}
\bibitem{Anguelova:2017djf}
L.~Anguelova, P.~Suranyi and L.~C.~R.~Wijewardhana,
% "Systematics of Constant Roll Inflation,"
JCAP \textbf{02}, 004 (2018)
doi:10.1088/1475-7516/2018/02/004
[arXiv:1710.06989 [hep-th]].

%\cite{Yi:2017mxs}
\bibitem{Yi:2017mxs}
Z.~Yi and Y.~Gong,
% "On the constant-roll inflation,"
JCAP \textbf{03}, 052 (2018)
doi:10.1088/1475-7516/2018/03/052
[arXiv:1712.07478 [gr-qc]].

%\cite{Gao:2018tdb}
\bibitem{Gao:2018tdb}
Q.~Gao, Y.~Gong and Q.~Fei,
% "Constant-roll tachyon inflation and observational constraints,"
JCAP \textbf{05}, 005 (2018)
doi:10.1088/1475-7516/2018/05/005
[arXiv:1801.09208 [gr-qc]].

%\cite{Randall:1999vf}
\bibitem{Randall:1999vf}
L.~Randall and R.~Sundrum,
% "An Alternative to Compactification,"
Phys. Rev. Lett. \textbf{83} (1999), 4690-4693
doi:10.1103/PhysRevLett.83.4690
[arXiv:hep-th/9906064 [hep-th]].
%
\bibitem{garousi}
M.~R.~Garousi,
%``Tachyon couplings on nonBPS D-branes and Dirac-Born-Infeld action,''
Nucl. Phys. B \textbf{584}, 284-299 (2000)
doi:10.1016/S0550-3213(00)00361-8
[arXiv:hep-th/0003122 [hep-th]].
%\cite{Steer:2003yu}
\bibitem{Steer:2003yu}
D.~A.~Steer and F.~Vernizzi,
% "Tachyon inflation: Tests and comparison with single scalar field inflation,"
Phys. Rev. D \textbf{70}, 043527 (2004)
doi:10.1103/PhysRevD.70.043527 
[arXiv:hep-th/0310139 [hep-th]].

%\cite{Liddle:1994dx}
\bibitem{Liddle:1994dx}
A.~R.~Liddle, P.~Parsons and J.~D.~Barrow,
% "Formalizing the slow roll approximation in inflation,"
Phys. Rev. D \textbf{50} (1994), 7222-7232
doi:10.1103/PhysRevD.50.7222
[arXiv:astro-ph/9408015 [astro-ph]].


%\cite{Martin:2012pe}
\bibitem{Martin:2012pe}
J.~Martin, H.~Motohashi and T.~Suyama,
% "Ultra Slow-Roll Inflation and the non-Gaussianity Consistency Relation,"
Phys. Rev. D \textbf{87}, no.2, 023514 (2013)
doi:10.1103/PhysRevD.87.023514
[arXiv:1211.0083 [astro-ph.CO]].


%\cite{Stojanovic:2023qgm}
\bibitem{Stojanovic:2023qgm}
M.~Stojanovic, N.~Bilic, D.~D.~Dimitrijevic, G.~S.~Djordjevic and M.~Milosevic,
%``Tachyon constant-roll inflation in Randall\textendash{}Sundrum II cosmology,''
Int. J. Mod. Phys. A \textbf{38}, no.32, 2343003 (2023)
doi:10.1142/S0217751X23430030
[arXiv:2306.02423 [gr-qc]].

%\cite{Mohammadi:2020ftb}
\bibitem{Mohammadi:2020ftb}
A.~Mohammadi, T.~Golanbari, S.~Nasri and K.~Saaidi,
% "Constant-roll brane inflation,"
Phys. Rev. D \textbf{101}, no.12, 123537 (2020)
doi:10.1103/PhysRevD.101.123537
[arXiv:2004.12137 [gr-qc]].

%\cite{Mohammadi:2018oku}
\bibitem{Mohammadi:2018oku}
A.~Mohammadi, K.~Saaidi and T.~Golanbari,
% "Tachyon constant-roll inflation,"
Phys. Rev. D \textbf{97} (2018) no.8, 083006
doi:10.1103/PhysRevD.97.083006
[arXiv:1801.03487 [hep-ph]].

%\cite{Planck:2018jri}
\bibitem{Planck:2018jri}
Y.~Akrami \textit{et al.} [Planck],
% "Planck 2018 results. X. Constraints on inflation,"
Astron. Astrophys. \textbf{641}, A10 (2020)
doi:10.1051/0004-6361/201833887
[arXiv:1807.06211 [astro-ph.CO]].

%\cite{Bilic:2018uqx}
\bibitem{Bilic:2018uqx}
N.~Bilic, D.~D.~Dimitrijevic, G.~S.~Djordjevic, M.~Milosevic and M.~Stojanovic,
% "Tachyon inflation in the holographic braneworld,"
JCAP \textbf{08} (2019), 034
doi:10.1088/1475-7516/2019/08/034
[arXiv:1809.07216 [gr-qc]].

%\cite{Mukhanov:2005sc}
\bibitem{Mukhanov:2005sc}
V.~Mukhanov,
Physical Foundations of Cosmology,
Cambridge University Press, 2005,
ISBN 978-0-521-56398-7
doi:10.1017/CBO9780511790553

%\cite{Guo:2003nz}
\bibitem{Guo:2003nz}
Z.~K.~Guo, H.~S.~Zhang and Y.~Z.~Zhang,
% "Inflationary attractor in braneworld scenario,"
Phys. Rev. D \textbf{69}, 063502 (2004)
doi:10.1103/PhysRevD.69.063502
[arXiv:hep-ph/0309163 [hep-ph]].
%

%\cite{Mohammadi:2022vru}
\bibitem{Mohammadi:2022vru}
A.~Mohammadi,
% "Constant-roll inflation driven by holographic dark energy,"
Phys. Dark Univ. \textbf{36}, 101055 (2022)
doi:10.1016/j.dark.2022.101055
[arXiv:2203.06643 [gr-qc]].

%\cite{Guo:2003zf}
\bibitem{Guo:2003zf}
Z.~K.~Guo, Y.~S.~Piao, R.~G.~Cai and Y.~Z.~Zhang,
% "Inflationary attractor from tachyonic matter,"
Phys. Rev. D \textbf{68}, 043508 (2003)
doi:10.1103/PhysRevD.68.043508
[arXiv:hep-ph/0304236 [hep-ph]].



\end{thebibliography}
\end{document}